\documentstyle[prd,preprint,tighten,aps,aps10,aps12,eqsecnum,amssymb,%
newlfont,prabib,epsfig,graphicx]{revtex}

\newcommand{\kkk}{$K^0 \bar K^0 \;$}
\newcommand{\cp}{$\mathcal{CP}\;$}
\newcommand{\ba}{\begin{array}}
\newcommand{\ea}{\end{array}}

\normalsize
\begin{document}

\preprint{\begin{tabular}{r}
UWThPh-2002-23\\
July 2002\\
\end{tabular}}
\title{Decoherence of entangled kaons and its connection to entanglement measures}


\author{Reinhold A. Bertlmann, Katharina Durstberger\\
and\\ Beatrix C. Hiesmayr\footnote{hiesmayr@ap.univie.ac.at}\\ \small{{\em Institute
for Theoretical Physics, University of Vienna}}\\ \small{{\em Boltzmanngasse 5, A-1090
Vienna, Austria}}}

\maketitle

\vfill

\begin{abstract}
We study the time evolution of the entangled kaon system by considering the Liouville -
von Neumann equation with an additional term which allows for decoherence. We choose as
generators of decoherence the projectors to the 2-particle eigenstates of the
Hamiltonian. Then we compare this model with the data of the CPLEAR experiment and find
in this way an upper bound on the strength $\lambda$ of the decoherence. We also relate
$\lambda$ to an effective decoherence parameter $\zeta$ considered previously in
literature. Finally we discuss our model in the light of different measures of
entanglement, i.e. the von Neumann entropy $S$, the entanglement of formation $E$ and the
concurrence $C$, and we relate the decoherence parameter $\zeta$ to the loss of
entanglement: $1 - E$.

\vspace{1.0cm}
\end{abstract}

\newpage

\section{Introduction}

Particle physics has become an interesting testing ground for fundamental questions of
quantum mechanics (QM). For instance, QM versus local realistic theories
\cite{Six,Selleri,PriviteraSelleri,BramonGarbarino2} and Bell inequalities
\cite{Ghirardi1,Uchiyama,BGH3,BH,BramonNowakowski,Ancochea,BramonGarbarino,Genovese} have
been tested. Furthermore, possible deviations from the quantum mechanical time evolution
have been studied, particularly in the neutral K-meson system
\cite{ellisHNS,ellisLMN,banksSP,huetP,BGH2,Benatti1,AndrianovTT} and B-meson system
\cite{DattaHome,BG97,dass,BG98,PompiliSelleri}. Recently also neutrino oscillations have
become of interest in this connection \cite{LisiMM}.

In this paper we concentrate on possible decoherence effects arising due to some
interaction of the system with its ``environment''. Sources for ``standard'' decoherence
effects are the strong interaction scatterings of kaons with nucleons, the weak
interaction decays and the noise of the experimental setup. ``Nonstandard'' decoherence
effects result from a fundamental modification of QM and can be traced back to the
influence of quantum gravity \cite{Hawking,tHooft1,tHooft2} -- quantum fluctuations in
the space-time structure on the Planck mass scale -- or to dynamical state reduction
theories \cite{GRW,pearle,gisinP,penrose}, and arise on a different energy scale.
However, we do not pursue further the reasons for decoherence effects, rather we want to
develop a specific model of decoherence and quantify the strength $\lambda$ of such
possible effects with the help of data of existing experiments.

For our model we focus on entangled massive particles moving apart in their center of
mass system, in particular on the \kkk system, where the strangeness $S=+,-$ plays the
role of spin ``up'' and ``down'' (for details see Ref. \cite{PHDtrixi}). We consider here
the famous EPR-like (Einstein, Podolsky, Rosen) scenario, as described by Bell
\cite{bell} for spin-$1/2$ particles, where the initial spin singlet state evolves in
time and after macroscopic separation the strangeness of the left and right moving
particle is measured. In contrast to other concepts in the literature we introduce
decoherence in the time evolution of the 2-particle entangled state which becomes
stronger with increasing distance between the two particles, whereas for the 1-particle
state we assume the usual quantum mechanical time evolution.

Then we compare our model of decoherence with the experimental data of the CPLEAR
experiment performed at CERN \cite{CPLEAR-EPR} and find an upper bound on possible
decoherence. We also can relate our model to an effective decoherence parameter $\zeta$,
introduced previously in literature, which quantifies the spontaneous factorization of
the wavefunction into product states (Furry--Schr\"odinger hypothesis \cite{Furry,Schrodinger}).

Finally we discuss our model within concepts of quantum information, where entanglement
is quantified by certain measures. We can connect directly the amount of decoherence of
the \kkk system parametrized by $\lambda$ or $\zeta$ with the loss of entanglement
expressed in terms of the concurrence or in terms of entanglement of formation. The
numerics for the information loss, the von Neumann entropy, and the entanglement loss of
the evolving \kkk system we illustrate in Fig.\ref{entropyentanglementfigure}.

\section{The model}

Let us begin our decoherence discussion with the 1-particle kaon system as an introduction.
Then we proceed to the case of two entangled neutral kaons and compare it with experimental
data.

\subsection{The 1-particle case}

We discuss the model of decoherence in a 2-dimensional Hilbert space
$\mathcal{H}=\mathbf{C}^2$ and consider the usual non-Hermitian ``effective mass''
Hamiltonian $H$ which describes the decay properties and the strangeness oscillations of
the kaons. The mass eigenstates, the short lived $|K_S\rangle$ and long lived
$|K_L\rangle$ states, are determined by
\begin{eqnarray}\label{Hamiltonian}
&&H\;|K_{S,L}\rangle\;=\;\lambda_{S,L}\;|K_{S,L}\rangle\qquad\textrm{with}\quad
\lambda_{S,L}\;=\;m_{S,L}-\frac{i}{2}\Gamma_{S,L} \, ,
\end{eqnarray}
with $m_{S,L}$ and $\Gamma_{S,L}$ being the corresponding masses and decay widths. For our
purpose \cp invariance\footnote{Note that corrections due to \cp violations are of order
$10^{-3}$, however, we compare this model of decoherence with the data of the CPLEAR
experiment \cite{CPLEAR-EPR} which are not sensitive to \cp violating effects.} is assumed,
i.e. the \cp eigenstates $|K_1^0\rangle , |K_2^0\rangle$ are equal to the mass eigenstates
\begin{eqnarray}\label{CPinvariance}
|K_1^0\rangle\equiv|K_S\rangle, \quad |K_2^0\rangle\equiv|K_L\rangle , \qquad\textrm{and}
\quad \langle K_S|K_L\rangle=0 \;.
\end{eqnarray}

As a starting point for our model of decoherence we consider the Liouville - von Neumann
equation with the Hamiltonian (\ref{Hamiltonian}) and allow for decoherence by adding a
so-called \textit{dissipator} $D[\rho]$, so that the time evolution of the density matrix $\rho$ is
governed by a master equation of the form
\begin{eqnarray}\label{Lindbladequation}
\frac{d\rho}{dt}&=&-i H\rho+i\rho H^\dagger-D[\rho] \; .
\end{eqnarray}
For the term $D[\rho]$ we choose the following ansatz (as in Ref. \cite{BG3})
\begin{eqnarray}\label{Dissipationterm}
D[\rho] \;=\; \lambda \, \big(P_S \rho P_L + P_L \rho P_S\big) \;=\;
\frac{\lambda}{2} \sum_{j=S,L} \big[P_j,[P_j,\rho]\big] \; ,
\end{eqnarray}
where $P_j\;=\;|K_j\rangle\langle K_j|$ ($j=S,L$) represent the projectors to the
eigenstates of the Hamiltonian and the \textit{decoherence parameter} $\lambda$ is
positive, $\lambda \ge 0$.

Apart from its simplicity ansatz (\ref{Dissipationterm}) has the
following nice features:
\begin{enumerate}
\item[i)] It generates a completely positive map since it is a
special case of Lindblad's general structure \cite{Lindblad}
\begin{eqnarray}
D[\rho] \;=\; \frac{1}{2} \sum_j (A^\dagger_j A_j\; \rho+\rho A^\dagger_j A_j-2 A_j
\rho A_j^\dagger)
\end{eqnarray}
if we identify $A_j = \sqrt{\lambda} \, P_j \; , \; j=S,L$.

Equivalently, it is a special form of the Gorini-Kossakowski-Sudarshan \cite{Sudarshan}
expression (see, e.g., Ref. \cite{BG4}).
\item[ii)] It conserves energy in case of a Hermitian Hamiltonian
since $[P_j,H]=0$ (see, e.g. Ref.~\cite{Adler1}).
\item[iii)] The von Neumann entropy $S(\rho)=-\mathrm{Tr}(\rho\ln\rho)$ is not
decreasing as a function of time since $P_j^\dagger = P_j$, thus $A_j^\dagger = A_j$
in our case, which is a theorem due to Narnhofer and Benatti \cite{BenattiNarnhofer}.
\end{enumerate}

With choice (\ref{Dissipationterm}) the time evolution (\ref{Lindbladequation}) decouples
for the components of $\rho$ which are defined by
\begin{equation}
\rho (t) \;= \sum_{i,j=S,L} \rho_{ij} (t) \, | K_i \rangle \langle K_j | \,,
\end{equation}
and we obtain
\begin{eqnarray}\label{lambdatimeevolution}
\rho_{SS}(t)&=&\rho_{SS}(0)\cdot e^{-\Gamma_S t}\nonumber\\
\rho_{LL}(t)&=&\rho_{LL}(0)\cdot e^{-\Gamma_L t}\nonumber\\
\rho_{LS}(t)&=&\rho_{LS}(0)\cdot e^{-i \Delta m t - \Gamma t - \lambda t}
\end{eqnarray}
with $\Delta m=m_L-m_S$ and $\Gamma=\frac{1}{2}(\Gamma_S+\Gamma_L)$.
Only the off-diagonal elements are effected by our model of decoherence.
Before discussing further the model we now proceed to the 2-particle system.

\subsection{The 2-particle case}

In the case of two entangled neutral kaons we make the following identification
\begin{eqnarray}\label{2-states}
|e_1\rangle\;=\;|K_S\rangle_l\otimes|K_L\rangle_r\qquad\textrm{and}\qquad
|e_2\rangle\;=\;|K_L\rangle_l\otimes|K_S\rangle_r \;,
\end{eqnarray}
and we consider -- as common -- the total Hamiltonian as a tensor product of the
1-particle Hilbert spaces: $H=H_l\otimes\mathbf{1}_r + \mathbf{1}_l\otimes H_r$, where
$l$ denotes the left-moving and $r$ the right-moving particle. Then the initial quasispin
singlet state
\begin{eqnarray}\label{singletstate}
|\psi^{-}\rangle&=&\frac{1}{\sqrt{2}}\biggl\lbrace |e_1\rangle-|e_2\rangle\biggr\rbrace \;,
\end{eqnarray}
is equivalently given by the density matrix
\begin{eqnarray}\label{rhozero}
\rho(0)\;=\;|\psi^{-}\rangle\langle \psi^{-}|\;=\;\frac{1}{2}\biggl\lbrace |e_1\rangle\langle e_1|
+|e_2\rangle\langle e_2|-|e_1\rangle\langle e_2|-|e_2\rangle\langle e_1|\biggr\rbrace\;.
\end{eqnarray}
Following the considerations of the 1-particle case we find that the time evolution given
by (\ref{Lindbladequation}) with our choice (\ref{Dissipationterm}), where now the
operators $P_j\;=\;|e_j\rangle\langle e_j|$ ($j=1,2$) project to the eigenstates of the
2-particle Hamiltonian $H$, also decouples
\begin{eqnarray}
\rho_{11}(t)&=&\rho_{11}(0)\; e^{-2 \Gamma t}\nonumber\\
\rho_{22}(t)&=&\rho_{22}(0)\; e^{-2 \Gamma t}\nonumber\\
\rho_{12}(t)&=&\rho_{12}(0)\, e^{-2 \Gamma t-\lambda t} \;.
\end{eqnarray}
Consequently we obtain for the time-dependent density matrix
\begin{eqnarray}\label{matrixevolutionsolution}
\rho(t) \;=\; \frac{1}{2} e^{-2 \Gamma t} \biggl\lbrace  |e_1\rangle\langle e_1| +
|e_2\rangle\langle e_2| - e^{-\lambda t} \big( \,|e_1\rangle\langle e_2| +
|e_2\rangle\langle e_1| \, \big) \biggr\rbrace \;.
\end{eqnarray}
The decoherence arises through the factor $e^{-\lambda t}$ which only effects the
off-diagonal elements. It means that for $t>0$ and $\lambda \not = 0$ the density matrix
$\rho(t)$ does not correspond to a pure state anymore (for further discussion, see
Section \ref{visualization}).

Note that the assumption of \cp invariance, Eq.(\ref{CPinvariance}), -- which is
sufficient for our purpose -- implying $\langle e_1|e_2\rangle = 0$, is crucial. Otherwise
we would have a time evolution into the full 4-dimensional Hilbert space of states.

\subsection{Bounds from experimental data}\label{connectiontoexperiment}

In order to obtain information on possible values of $\lambda$ we compare our model
of decoherence with data of the CPLEAR experiment performed at CERN \cite{CPLEAR-EPR}.
We have the following point of view. The 2-particle density matrix follows
the time evolution given by Eq.(\ref{Lindbladequation}) with the Lindblad generators
$A_j=\sqrt{\lambda}\; |e_j\rangle\langle e_j|$ and undergoes thereby some decoherence.
We measure the strangeness content $S$ of the left-moving particle at time $t_l$ and
of the right-moving particle at time $t_r$. For sake of definiteness we choose
$t_r\leq t_l$. For times $t_r\leq t\leq t_l$ we have a 1-particle state which evolves
exactly according to QM, i.e. no further decoherence is picked up.

Mathematically, the measurement of the strangeness content, i.e. the right-moving
particle being a $K^0$ or a $\bar K^0$ at time $t_r$, is obtained by
\begin{eqnarray}
\mathrm{Tr}_r \{ \mathbf{1}_l \otimes |S^{'}\rangle\langle S^{'}|_r \;\; \rho(t_r) \}
&\equiv& \rho_l(t=t_r;t_r) \;,
\end{eqnarray}
with strangeness $S^{'} = +,-$ and $|+\rangle = |K^0\rangle$, $|-\rangle = |\bar
K^0\rangle$. Consequently, $\rho_l(t;t_r)$ for times $t \ge t_r$ is the 1-particle
density matrix for the left-moving particle and evolves as a 1-particle state according
to pure QM. At $t=t_l$ the strangeness content ($S=+,-$) of the second particle is
measured and we finally have the probability
\begin{eqnarray}
P_\lambda(S,t_l;S^{'},t_r)&=&\mathrm{Tr}_l\{|S\rangle\langle S|_l\;\;\rho_l(t_l;t_r)\}\;.
\end{eqnarray}
Explicitly, we find the following results for the like- and unlike-strangeness probabilities
\begin{eqnarray}\label{lambdaprobabilities}
\lefteqn{P_\lambda(K^0,t_l;K^0,t_r)\;=\;P_\lambda(\bar K^0,t_l;\bar K^0,t_r)\;=}\nonumber\\
&=&\frac{1}{8} \biggl\lbrace
e^{-\Gamma_S t_l-\Gamma_L t_r}+e^{-\Gamma_L t_l-\Gamma_S t_r}-e^{-\lambda t_r}\;2
\cos(\Delta m \Delta t)\cdot e^{-\Gamma (t_l+t_r)}\biggr\rbrace\nonumber\\
\lefteqn{P_\lambda(K^0,t_l;\bar K^0,t_r)\;=\;P_\lambda(\bar K^0,t_l;K^0,t_r)\;=}\nonumber\\
&=&\frac{1}{8} \biggl\lbrace
e^{-\Gamma_S t_l-\Gamma_L t_r}+e^{-\Gamma_L t_l-\Gamma_S t_r}+e^{-\lambda t_r}\;2
\cos(\Delta m \Delta t)\cdot e^{-\Gamma (t_l+t_r)}\biggr\rbrace \;,
\end{eqnarray}
with $\Delta t = t_l-t_r$.

Note that at equal times $t_l=t_r=t$ the like-strangeness probabilities
\begin{equation}
P_\lambda(K^0,t;K^0,t)\;=\;P_\lambda(\bar K^0,t;\bar K^0,t)\;=\;
\frac{1}{4} \; e^{-2\Gamma t} \; (1 - e^{-\lambda t})
\end{equation}
do not vanish, in contrast to the quantum mechanical EPR-correlation.\\

The asymmetry of the probabilities is directly sensitive to the interference term and
has been measured by the CPLEAR collaboration \cite{CPLEAR-EPR}. For pure QM we have
\begin{eqnarray}\label{qmasymmetry}
\lefteqn{A^{QM}(\Delta t) =} \nonumber\\
&=& \frac{P(K^0,t_l;\bar K^0,t_r)+P(\bar K^0,t_l;K^0,t_r)-P(K^0,t_l;K^0,t_r)
-P(\bar K^0,t_l;\bar K^0,t_r)}
{P(K^0,t_l;\bar K^0,t_r)+P(\bar K^0,t_l; K^0, t_r)
+P(K^0,t_l;K^0,t_r)+P(\bar K^0,t_l;\bar K^0,t_r)}\nonumber\\
&=& \;\frac{\cos(\Delta m \Delta t)}{\cosh(\frac{1}{2}\Delta \Gamma \Delta t)} \;,
\end{eqnarray}
with $\Delta\Gamma = \Gamma_L-\Gamma_S$, and for our decoherence model we find
by inserting the probabilities (\ref{lambdaprobabilities})
\begin{eqnarray}\label{lambdaasymmetry}
A^\lambda(t_l,t_r)&=&\frac{\cos(\Delta m \Delta t)}{\cosh(\frac{1}{2}\Delta \Gamma\Delta t)}
\cdot e^{-\lambda \min{\{t_l,t_r\}}}\;=\;
A^{QM}(\Delta t)\cdot e^{-\lambda \min{\{t_l,t_r\}}} \;.
\end{eqnarray}
Thus the decoherence effect, simply given by the factor $e^{-\lambda \min{\{t_l,t_r\}}}$,
depends only -- due to our philosophy -- on the time of the first measured kaon, in our case:
$\min{\{t_l,t_r\}} = t_r$.\\

Comparing now our model with the results of the CPLEAR experiment \cite{CPLEAR-EPR}
we recall that the experimental set-up has two configurations:
In the first configuration both kaons propagate $2cm$, in the second one kaon propagates
$2cm$ and the other kaon $7cm$ until they are measured by an absorber.

Fitting the decoherence parameter $\lambda$ by comparing the asymmetry
(\ref{lambdaasymmetry}) with the experimental data \cite{CPLEAR-EPR} we find, when
averaging over both configurations\footnote{We have scaled $\Delta t$ in the QM asymmetry
(\ref{qmasymmetry}) in order to reproduce the QM curve in Fig. 9 of the CPLEAR
collaboration \cite{CPLEAR-EPR}.}, the following bounds on $\lambda$
\begin{eqnarray}\label{ergebnis}
\bar\lambda \;=\; (1.84^{+2.50}_{-2.17})\cdot 10^{-12}\;\textrm{MeV} \quad\textrm{and}\quad
\bar\Lambda \;=\; \frac{\bar\lambda}{\Gamma_S}\;=\;0.25^{+0.34}_{-0.32}\;.
\end{eqnarray}
The results (\ref{ergebnis}) are certainly compatible with QM ($\lambda = 0$), nevertheless,
the experimental data allow an upper bound
$\bar\lambda_{up} = 4.34 \cdot 10^{-12}\;\textrm{MeV}$ for possible decoherence in the
entangled \kkk $\,$ system.

The results (\ref{ergebnis}) can be compared with the analogous investigation of the
entangled $B^0 \bar B^0$ system \cite{BG3}, which gives
$\lambda_B = (-47\pm76)\cdot 10^{-12}\;\textrm{MeV}$. Thus we find that the bounds
(\ref{ergebnis}) of the \kkk system are an order of magnitude more restrictive.

\section{Connection to a phenomenological model}

There exists a one-to-one correspondence between the model of decoherence
(\ref{Lindbladequation}) and a phenomenological model \cite{dass,BGH1,H1,Eberhard2} where
the decoherence is introduced by multiplying the interference term of the transition
amplitude by the factor $(1-\zeta)$. The quantity $\zeta$ is called the
\textit{effective decoherence parameter}. Our initial state is again the spin singlet state
$|\psi^- \rangle$ which is given by the mass eigenstate representation
(\ref{singletstate}) then we get for the like-strangeness probability
\begin{eqnarray}\label{plikezeta}
\lefteqn{P(K^0,t_l;K^0,t_r)\;=\;||\langle K^0|_l \otimes \langle K^0|_r \;
|\psi^-(t_l,t_r)\rangle||^2 \quad\longrightarrow\quad P_{\zeta}(K^0, t_l; K^0, t_r)=} \nonumber\\
&=&\frac{1}{2}\biggl\lbrace e^{-\Gamma_S t_l-\Gamma_L t_r} |\langle K^0|K_S\rangle_l|^2\;
|\langle K^0|K_L\rangle_r|^2 + e^{-\Gamma_L t_l-\Gamma_S t_r}|\langle K^0|K_L\rangle_l|^2\;
|\langle K^0|K_S\rangle_r|^2 \nonumber\\
& &-2 \underbrace{(1-\zeta)}\; \mathrm{Re}\{\langle K^0|K_S\rangle_l^*
\langle K^0|K_L\rangle_r^* \langle K^0|K_L\rangle_l \langle K^0|K_S\rangle_r\;
e^{-i \Delta m \Delta t}\} \cdot e^{-\Gamma(t_l+t_r)} \biggr\rbrace \nonumber\\
& &\quad\textrm{modification} \nonumber\\
&=&\frac{1}{8}\biggl\lbrace e^{-\Gamma_S t_l-\Gamma_L t_r} + e^{-\Gamma_L t_l-\Gamma_S t_r}
-2 \underbrace{(1-\zeta)} \cos(\Delta m \Delta t) \cdot e^{-\Gamma(t_l+t_r)}\biggr\rbrace\;,
\nonumber\\
& &\hphantom{\frac{1}{8}\biggl\lbrace e^{-\Gamma_S t_l-\Gamma_L t_r} +
e^{-\Gamma_L t_l-\Gamma_S t_r}} \quad\textrm{modification}
\end{eqnarray}
and the unlike-strangeness probability just changes the sign of the interference term.

The value $\zeta=0$ corresponds to pure QM and $\zeta=1$ to total decoherence
or spontaneous factorization of the wave function, which is commonly known as
Furry--Schr\"odinger hypothesis \cite{Furry,Schrodinger}.
The effective decoherence parameter $\zeta$, introduced in this way ``by hand'', interpolates
continuously between these two limits and represents a measure for the amount of
decoherence which results in a loss of entanglement of the total quantum state
(we come back to this point in Section \ref{visualization}).\\

Calculating the asymmetry of the strangeness events with the probabilities (\ref{plikezeta})
we obtain
\begin{eqnarray}\label{zetaasymmetry}
A^{\zeta}(t_l,t_r) \;=\; A^{QM}(\Delta t) \cdot \big(\,1-\zeta(t_l,t_r)\,\big)\;.
\end{eqnarray}
When we compare now the two approaches, i.e. Eq.(\ref{lambdaasymmetry}) with
Eq.(\ref{zetaasymmetry}), we find the formula
\begin{equation}\label{zetamin}
\zeta(t_l,t_r) \;=\; 1 - e^{-\lambda \min{\{t_l,t_r\}}} \;.
\end{equation}
Of course, the values (\ref{ergebnis}) are in agreement with the corresponding $\zeta$
values (averaged over both experimental setups): $\bar\zeta\,=\,0.13^{+0.16}_{-0.15}\,$,
as derived in Refs. \cite{BGH1,MTtrixi}.

We consider the decoherence parameter $\lambda$ to be the fundamental constant, whereas
the value of the effective decoherence parameter $\zeta$ depends on the time when a
measurement is performed. In the time evolution of the state $|\psi^- \rangle$,
Eq.(\ref{singletstate}), represented by the density matrix (\ref{matrixevolutionsolution}),
we have the relation
\begin{equation}\label{zeta}
\zeta(t) \;=\; 1 - e^{-\lambda t} \;,
\end{equation}
which after the measurement of the left- and right moving particles at $t_l$ and $t_r$
turns into formula (\ref{zetamin}), when decoherence is implemented at the 2-particle
level.\\

Our model can be compared with the case where decoherence in the time evolution
(\ref{Lindbladequation}) happens at a 1-particle level and is transferred to the 2-particle
level by a tensor product of the 1-particle Hilbert spaces \cite{Benatti1}. Using
the same structure of the decoherence term (\ref{Dissipationterm}), where now the operators
project to the 1-particle states instead of states (\ref{2-states}), we obtain the relation
\begin{equation}\label{zeta-lr}
\zeta(t_l,t_r) \;=\; 1 - e^{-\lambda \,(t_l+t_r)} \;.
\end{equation}
Here the parameter $\zeta$ depends explicitly on both times $t_l$ and $t_r$, the
``eigentimes'' of the left- and right-moving kaon, instead of one time $\min{\{t_l,t_r\}}$,
the time of the system when the first kaon is measured.

Measuring the strangeness content of the entangled kaons at definite times
we have the possibility to distinguish experimentally these two models (\ref{zetamin}) and
(\ref{zeta-lr}) on the basis of time dependent event rates. Indeed, it would be of high
interest to measure in future experiments the asymmetry of the strangeness events for
several different times, in order to confirm the time dependence of the
decoherence effect.
In fact, such a possibility is now offered in the B-meson system. Entangled
$B^0\bar B^0$ pairs are created with high density at the asymmetric B-factories and
identified by the detectors BELLE at KEK-B (see e.g. Refs. \cite{Belle,Leder}) and BABAR
at PEP-II (see e.g. Refs. \cite{Aubert,Babar}) with a high resolution at different
distances or times.

\section{Decoherence and entanglement loss}\label{visualization}

The term $D[\rho]$ in the master equation (\ref{Lindbladequation}) is usually called
dissipative term or dissipator (see, e.g., Ref. \cite{BreuerPetruccione}).
In general, $D[\rho]$ describes 2 phenomena occurring in an open quantum system $S$,
namely decoherence and dissipation. When the system $S$ interacts with the environment
$E$ the initially product state evolves into entangled states of $S+E$ in the course of
time \cite{Joos,KublerZeh}. It leads to mixed states in $S$ -- what means decoherence --
and to an energy exchange between $S$ and $E$ -- what is called dissipation.

The decoherence destroys the occurrence of longrange quantum correlations by suppressing
the off-diagonal elements of the density matrix in a given basis and leads to an
information transfer from $S$ to $E$.

In general, both effects are present, however, decoherence acts on a much shorter time
scale \cite{Joos,JoosZeh,Zurek,Alicki} than dissipation and is the more important effect
in quantum information processes.

Our model describes decoherence and not dissipation. The increase of decoherence of the
initially totally entangled \kkk system as time evolves means on the other hand a
decrease of entanglement of the system. This loss of entanglement we can quantify and
visualize explicitly.

In the field of quantum information the entanglement of a state is quantified by introducing
certain measures. In this connection the entropy plays a fundamental role, which measures
``somehow'' the degree of uncertainty of a quantum state. A common measure is the von Neumann
entropy function $S$, the entanglement of formation $E$ and the concurrence $C$.

\subsection{Von Neumann entropy}

Since we are only interested in the effect of decoherence, we want to properly normalize
the state (\ref{matrixevolutionsolution}) in order to compensate the decay
property\footnote{Note that for other physical situations -- like the verification of
Bell inequalities -- the decay property must not be neglected \cite{BH}.} of the
non-Hermitian Hamiltonian $H$
\begin{eqnarray}\label{normdensitymatrix}
\rho_N(t) &=& \frac{\rho(t)}{\mathrm{Tr}\rho(t)} \;.
\end{eqnarray}
Then von Neumann's entropy function for the state (\ref{normdensitymatrix})
gives
\begin{eqnarray}\label{vonNeumannentropy}
S\big(\rho_N(t)\big) &=& -\mathrm{Tr}\{\rho_N(t)\log_2 \rho_N(t)\}\nonumber\\
&=& - \frac{1-e^{-\lambda t}}{2}\log_2 \frac{1-e^{-\lambda t}}{2} -
\frac{1+e^{-\lambda t}}{2}\log_2 \frac{1+e^{-\lambda t}}{2} \;.
\end{eqnarray}
At the time $t=0$ the entropy is zero, there is no uncertainty in the system, the quantum
state is pure and maximally entangled. For $t>0$ the entropy gets nonzero, increases and
approaches the value $1$ for $t \to \infty$. Hence the state becomes more and more mixed.
Mixed states provide only partial information about the system, and the entropy measures
how much of the maximal information is missing (see, e.g., Ref. \cite{Thirring}).
In Fig.\ref{entropyentanglementfigure} the von Neumann entropy $S(\rho_N(t))$ is plotted for
the mean value and upper bound of the decoherence parameter $\lambda$, Eq.(\ref{ergebnis}),
as determined from the CPLEAR experiment \cite{CPLEAR-EPR}.

Let us consider next the reduced density matrices of the subsystems, i.e. the propagating kaons
on the left $l$ and right $r$ hand side
\begin{equation}\label{reduceddensity}
\rho^{\,l}_N(t) = \mathrm{Tr}_r\{\rho_N(t)\} \quad \textrm{and} \quad
\rho^{\,r}_N(t) = \mathrm{Tr}_l\{\rho_N(t)\} \;.
\end{equation}
Then the uncertainty in the subsystem $l$ before the subsystem $r$ is measured is
given by the von Neumann entropy $S(\rho^{\,l}_N(t))$ of the corresponding reduced
density matrix $\rho^{\,l}_N(t)$ (and alternatively we can replace $l \to r$). In our case
we find
\begin{equation}\label{entropysubsystems}
S\big(\rho^{\,l}_N(t)\big) = S\big(\rho^{\,r}_N(t)\big) = 1  \quad \forall \; t \geq 0 \;.
\end{equation}
We see that the reduced entropies are independent of $\lambda$. The correlation stored in
the composite system is, with increasing time, lost into the environment -- what is
expected intuitively -- and {\em not} into the subsystems, i.e. the individual kaons.

Note that because of our chosen normalization (\ref{normdensitymatrix})
the effects seen are only due to the introduced decoherence via the term $D[\rho]$
(\ref{Dissipationterm}) and not due to the decay of the system.

\subsection{Lack of separability}

For the subsequent considerations it is convenient to recall the ``quasispin'' picture
for the \kkk system (see, e.g., Ref. \cite{BH}) in order to express the formulae in terms
of Pauli spin matrices. Then the projection operators to the mass eigenstates represent
the spin projection operators ``up'' and ``down''
\begin{eqnarray}
P_S &=& |K_S\rangle\langle K_S| \;=\; \sigma_{\uparrow} \;=\; \frac{1}{2} \,
(\mathbf{1} + \sigma_z) \;=\;
\left( \ba{cc}
1 & 0\\
0 & 0\\
\ea \right) \;,\nonumber\\
P_L &=& |K_L\rangle\langle K_L| \;=\; \sigma_{\downarrow} \;=\; \frac{1}{2} \,
(\mathbf{1} - \sigma_z) \;=\;
\left( \ba{cc}
0 & 0\\
0 & 1\\
\ea \right) \;,
\end{eqnarray}
and the transition operators are the ``spin-ladder'' operators
\begin{eqnarray}
P_{SL} &=& |K_S\rangle\langle K_L| \;=\; \sigma_{+} \;=\; \frac{1}{2} \,
(\sigma_x + i \, \sigma_y) \;=\;
\left( \ba{cc}
0 & 1\\
0 & 0\\
\ea \right) \;,\nonumber\\
P_{LS} &=& |K_L\rangle\langle K_S| \;=\; \sigma_{-} \;=\; \frac{1}{2} \,
(\sigma_x - i \, \sigma_y) \;=\;
\left( \ba{cc}
0 & 0\\
1 & 0\\
\ea \right) \;.
\end{eqnarray}
With help of these spin matrices the density matrix (\ref{normdensitymatrix})
can be expressed by
\begin{equation}\label{densitymatrixspin}
\rho_N(t) \,=\, \frac{1}{2}\big\{\sigma_{\uparrow}\otimes\sigma_{\downarrow}
+ \sigma_{\downarrow}\otimes\sigma_{\uparrow} \,-\, e^{-\lambda t}\,
[\sigma_{+}\otimes\sigma_{-} + \sigma_{-}\otimes\sigma_{+}]\big\} \;,
\end{equation}
or by
\begin{equation}\label{densitymatrixspinxyz}
\rho_N(t) \,=\, \frac{1}{4}\big\{\mathbf{1} - \sigma_z\otimes\sigma_z \,-\, e^{-\lambda t}\,
[\sigma_x\otimes\sigma_x + \sigma_y\otimes\sigma_y]\big\} \;,
\end{equation}
which at $t=0$ coincides with the well-known expression for the pure spin singlet state
$\rho_N(t=0) \,=\, \frac{1}{4} \, (\mathbf{1} - \vec\sigma\otimes\vec\sigma)$;
see, e.g., Ref. \cite{BNT}.
Operators (\ref{densitymatrixspin}),  (\ref{densitymatrixspinxyz}) can be nicely written as
$4 \times 4$ matrix
\begin{eqnarray}\label{densitymatrix4x4}
\rho_N(t) \;=\; \frac{1}{2}
\left( \ba{cccc}
0 & 0 & 0 & 0\\
0 & 1 & -e^{-\lambda t} & 0\\
0 & -e^{-\lambda t} & 1 & 0\\
0 & 0 & 0 & 0\\
\ea \right) \;.
\end{eqnarray}\\

It is also illustrative to choose for the density matrix $\rho_N(t)$
another basis, the socalled ``Bell basis''
\begin{eqnarray}
\rho^{\mp} \;=\; |\psi^{\mp}\rangle\langle \psi^{\mp}| \qquad\textrm{and}\qquad
\omega^{\mp} \;=\; |\phi^{\mp}\rangle\langle \phi^{\mp}| \;,
\end{eqnarray}
with $|\psi^{-}\rangle$ given by Eq.(\ref{singletstate}) and  $|\psi^{+}\rangle$ by
\begin{eqnarray}\label{symmetricstate}
|\psi^{+}\rangle&=&\frac{1}{\sqrt{2}}\biggl\lbrace |e_1\rangle+|e_2\rangle\biggr\rbrace \;.
\end{eqnarray}
The states $|\phi^{\mp}\rangle = \frac{1}{\sqrt{2}}
(|\uparrow\uparrow\rangle \mp |\downarrow\downarrow\rangle)$ (in spin notation) do not
contribute here. Then we are led to the following proposition.

\vspace{0.5cm}

\noindent{\bf Proposition:}
{\em The state represented by the density matrix $\rho_N(t)$ (\ref{normdensitymatrix})
becomes mixed for $0<t<\infty$ but remains entangled. Separability is achieved
asymptotically $t \to \infty$ with the weight $e^{-\lambda t}$. Explicitly, $\rho_N(t)$
is the following mixture of the Bell states $\rho^{-}$ and $\rho^{+}\,$}:
\begin{equation}\label{densitymatrixBell}
\rho_N(t) \;=\; \frac{1}{2} \big(1 + e^{-\lambda t}\big) \, \rho^{-} \;+\;
\frac{1}{2} \big(1 - e^{-\lambda t}\big) \, \rho^{+} \;.\\
\end{equation}

\noindent{\bf Proof:}
\begin{itemize}
\item[i)] Diagonalizing matrix (\ref{densitymatrix4x4}) we find
$\big\{\frac{1}{2}\big(1 + e^{-\lambda t}\big),
\frac{1}{2}\big(1 - e^{-\lambda t}\big),0,0\big\}$,
the eigenvalues together with the eigenvectors $\rho^{-}$ and $\rho^{+}\,$, which proves
Eq.(\ref{densitymatrixBell}).
For $t=0$ the matrix $\rho_N(0)$ is a projector to a 1-dimensional subspace: $\{1,0,0,0\}$,
the Bell state $\rho^{-}\,$, so the state is pure.
For $0<t<\infty$ the matrix $\rho_N(t)$ is no longer a projector, thus
the state is mixed. The state becomes asymptotically $t \to \infty$ totally mixed,
separable. Of course, $\rho_N(t)$ also satisfies the
\textit{mixed state criterion}
\begin{eqnarray}\label{densitymatrixsquared4x4}
\rho^{\,2}_N(t) \;=\; \frac{1}{4}
\left( \ba{cccc}
0 & 0 & 0 & 0\\
0 & 1+e^{-2\lambda t} & -2e^{-\lambda t} & 0\\
0 & -2e^{-\lambda t} & 1+e^{-2\lambda t} & 0\\
0 & 0 & 0 & 0\\
\ea \right)
\;\not=\; \rho_N(t) \quad \textrm{for} \quad t>0 \;.
\end{eqnarray}

\item[ii)] Entanglement we prove via lack of separability.
According to Peres \cite{Peres} and the Horodeckis \cite{Horodecki2} separability is
determined by the \textit{positive partial transposition criterion}: the partial
transposition of a separable state with respect to any subsystem is positive (i.e. the
operator has positive eigenvalues). Applying the transposition operator $T$, defined by
$T (\sigma^i)_{kl} = (\sigma^i)_{lk} \,$, to the left- or right hand side, we find a negative
eigenvalue:
\begin{eqnarray}
(\mathbf{1}_l \otimes T_r) \, \rho_N(t) \,&=&\,
\frac{1}{2}\big\{\sigma_{\uparrow}\otimes\sigma_{\downarrow} +
\sigma_{\downarrow}\otimes\sigma_{\uparrow} \,-\, e^{-\lambda t}\,
[\sigma_{+}\otimes\sigma_{+} + \sigma_{-}\otimes\sigma_{-}]\big\}\nonumber\\
\;&=&\; \frac{1}{2}
\left( \ba{cccc}
0 & 0 & 0 & -e^{-\lambda t}\\
0 & 1 & 0 & 0\\
0 & 0 & 1 & 0\\
-e^{-\lambda t} & 0 & 0 & 0\\
\ea \right) \not\geq 0 \;,
\end{eqnarray}
with eigenvalues $\big\{\frac{1}{2},\frac{1}{2},\frac{1}{2} e^{-\lambda t},
-\frac{1}{2} e^{-\lambda t}\big\}\;$.

Alternatively, we could use Horodecki's  \cite{Horodecki}
\textit{reduction criterion}: $\rho_N(t)$ is separable iff
$\rho^{\,l}_N(t)\otimes\mathbf{1}_r - \rho_N(t) \ge 0$ or
$\mathbf{1}_l\otimes\rho^{\,r}_N(t) - \rho_N(t) \ge 0$.
In our case we have
\begin{eqnarray}
\mathbf{1}_l\otimes\rho^{\,r}_N(t) - \rho_N(t) \;=\; \frac{1}{2}
\left( \ba{cccc}
1 & 0 & 0 & 0\\
0 & 0 & e^{-\lambda t} & 0\\
0 & e^{-\lambda t} & 0 & 0\\
0 & 0 & 0 & 1\\
\ea \right) \not\geq 0 \;,
\end{eqnarray}
where the eigenvalues are the same as for the previous criterion.
For a general separability criterion, a generalized Bell inequality, see Ref. \cite{BNT}.
$\Box$
\end{itemize}

\subsection{Entanglement of formation and concurrence}

For pure quantum states entanglement can be measured by the entropy of the reduced
density matrices. For mixed states, however, the von Neumann entropy is not generally a
good measure for entanglement, whereas another measure, \textit{entanglement of formation}
\cite{Bennett}, is very suitable. It has been constructed by the authors \cite{Bennett} to
quantify the resources needed to create a given entangled state.

\subsubsection{Definitions}

Every density matrix $\rho$ can be decomposed in an ensemble of pure states
$\rho_i=|\psi_i\rangle\langle\psi_i|$
with the probability $p_i$, i.e. $\rho=\sum_i p_i \rho_i$. The
entanglement of formation for a pure state is given by the entropy of either of the
two subsystems. For a mixed state the entanglement of formation for a bipartite system
is then defined as the average entanglement of the pure states of the decomposition,
minimized over all decompositions of $\rho$
\begin{eqnarray}\label{entanglementofformation}
E(\rho)&=&\min\sum_i \,p_i\, S(\rho_i^{\,l})\;.
\end{eqnarray}
Bennett et al. \cite{Bennett} found a remarkable simple formula for entanglement of
formation
\begin{eqnarray}\label{entanglementbound}
E(\rho) &\ge& \mathcal{E}\big(f(\rho)\big) \;,
\end{eqnarray}
where the function $\mathcal{E}\big(f(\rho)\big)$ is defined by
\begin{eqnarray}\label{entanglementformula-f}
\mathcal{E}\big(f(\rho)\big) \;=\;
H\bigg(\frac{1}{2} + \sqrt{f(1-f)}\bigg)
\quad \textrm{for} \quad f \ge \frac{1}{2} \;,
\end{eqnarray}
and $\mathcal{E}\big(f(\rho)\big) = 0$ for $f < \frac{1}{2}$.
The function $H$ represents the familiar binary entropy function
$H(x) = -x\log_2x - (1-x)\log_2(1-x)$.
The quantity $f$ is called the \textit{fully entangled fraction} of $\rho$
\begin{eqnarray}
f = \max \, \langle e|\rho|e \rangle \;,
\end{eqnarray}
being the maximum over all completely entangled states $|e \rangle$.
For general mixed states $\rho$ the function $\mathcal{E}\big(f(\rho)\big)$ is
only a lower bound to the entropy $E(\rho)$. For pure states and mixtures of Bell states
-- the case of our model -- the bound is saturated, $E=\mathcal{E}$, and we have
formula (\ref{entanglementformula-f}) for calculating the entanglement of formation.

Wootters and Hill \cite{Wootters1,Wootters2,Wootters3} found that
entanglement of formation for a general mixed state $\rho$ of two qubits can be
expressed by another quantity, the \textit{concurrence} $C$
\begin{eqnarray}\label{entanglementformula-C}
E(\rho) &=& \mathcal{E}\big(C(\rho)\big) \;=\;
H\bigg(\frac{1}{2} + \frac{1}{2}\sqrt{1-C^2}\bigg)
\quad \textrm{with} \quad 0 \le C \le 1 \;.
\end{eqnarray}
Explicitly, the function $\mathcal{E}(C)$ looks like
\begin{eqnarray}
\mathcal{E}(C)&=&-\frac{1+\sqrt{1-C^2}}{2}\log_2\frac{1+\sqrt{1-C^2}}{2}-
\frac{1-\sqrt{1-C^2}}{2}\log_2\frac{1-\sqrt{1-C^2}}{2}\nonumber\\
\end{eqnarray}
and is monotonically increasing from $0$ to $1$ as $C$ runs from $0$ to $1$.
Thus $C$ itself is a kind of entanglement measure in its own right.

Defining the spin flipped state $\tilde\rho$ of $\rho$ by
\begin{eqnarray}
\tilde\rho &=& (\sigma_y\otimes\sigma_y)\, \rho^* (\sigma_y\otimes\sigma_y) \;,
\end{eqnarray}
where $\rho^*$ is the complex conjugate and is taken in the standard basis, i.e. the basis
$\{|\uparrow\uparrow\rangle, |\downarrow\downarrow\rangle, |\uparrow\downarrow\rangle,
|\downarrow\uparrow\rangle\}$,
the concurrence $C$ is given by the formula
\begin{eqnarray}
C(\rho) &=& \max\{0,\lambda_1-\lambda_2-\lambda_3-\lambda_4\} \;.
\end{eqnarray}
The $\lambda_i$'s are the square roots of the eigenvalues, in decreasing order, of
the matrix $\rho\tilde\rho$.

\subsubsection{Applications to our model}

For the density matrix $\rho_N(t)$ (\ref{normdensitymatrix}) of our model, which is invariant
under spin flip (see, e.g., Eq.(\ref{densitymatrixspin})), i.e. $\tilde\rho_N = \rho_N$ and
thus $\rho_N\tilde\rho_N = \rho^{\,2}_N$, we get for the concurrence
\begin{eqnarray}
C\big(\rho_N(t)\big) &=& \max\big\{0,e^{-\lambda t}\big\} \;=\; e^{-\lambda t} \;,
\end{eqnarray}
and for the fully entangled fraction of $\rho_N(t)$
\begin{eqnarray}
f\big(\rho_N(t)\big) = \frac{1}{2}\big(1 + e^{-\lambda t}\big) \;,
\end{eqnarray}
which is simply the largest eigenvalue of $\rho_N(t)$. Clearly, in our case the
functions $C$ and $f$ are related by
\begin{eqnarray}
C\big(\rho_N(t)\big) &=& 2 \, f\big(\rho_N(t)\big) - 1 \;.
\end{eqnarray}
Finally, for the entanglement of formation of the \kkk system we have
\begin{eqnarray}\label{entanglementofformationforlambda}
E\big(\rho_N(t)\big)&=&-\frac{1+\sqrt{1-e^{-2\lambda t}}}{2}
\log_2\frac{1+\sqrt{1-e^{-2\lambda t}}}{2}\nonumber\\
& &-\frac{1-\sqrt{1-e^{-2\lambda t}}}{2}\log_2\frac{1-\sqrt{1-e^{-2\lambda t}}}{2}\;.
\end{eqnarray}
Recalling our relation (\ref{zeta}) between the decoherence parameters $\lambda$ and $\zeta$
we find a direct connection between the entanglement measure and the amount of decoherence
of the quantum system. Defining the \textit{loss of entanglement} as one minus entanglement
and expanding expression (\ref{entanglementofformationforlambda}) for small values of
$\lambda$ or $\zeta$ we obtain
\begin{eqnarray}
1 - C\big(\rho_N(t)\big) &=& \zeta(t)\label{entanglementlossC} \;,\\
1 - E\big(\rho_N(t)\big) &\doteq& \frac{1}{\ln2}\;\zeta(t) \;\doteq\;
\frac{\lambda}{\ln2}\;t\label{entanglementlossE} \;.
\end{eqnarray}

The loss of entanglement of the propagating \kkk system in terms of the concurrence $C$,
Eq.(\ref{entanglementlossC}), equals precisely the amount of the decoherence parameter $\zeta$
which describes the factorization of the initial spin singlet state into the product states
$|K_S\rangle_l\otimes|K_L\rangle_r$ or $|K_L\rangle_l\otimes|K_S\rangle_r$ (Furry--Schr\"odinger
hypothesis). In terms of entanglement of formation, Eq.(\ref{entanglementlossE}), the decoherence
parameter is weighted by a factor $1/\ln2 = 1.44$ (the factor $\ln2$ reflects the dimension
$2$ of the $K$-meson quasispin space) and the entanglement loss increases linearly with time.

\subsection{Discussion of the results}

\begin{figure}
\center{\includegraphics[width=310pt, height=180pt, keepaspectratio=true]{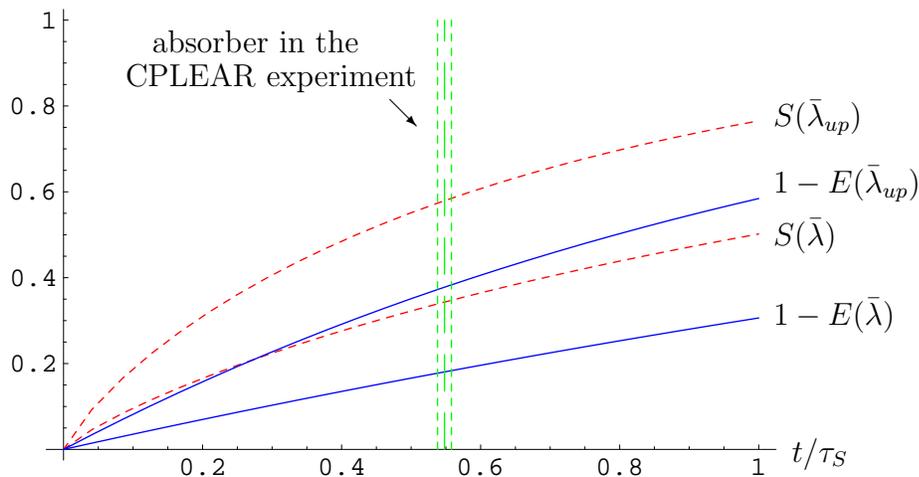}
\begin{picture}(1,1)(0,0)
\put(-240,163){absorber in the}
\put(-250,150){CPLEAR experiment}
\put(-150,145){\vector(1,-1){10}}
\put(-5,135){$S(\bar\lambda_{up})$}
\put(-5,110){$1-E(\bar\lambda_{up})$}
\put(-5,90){$S(\bar\lambda)$}
\put(-5,60){$1-E(\bar\lambda)$}
\put(2,8){$t/\tau_S$}
\end{picture}}
\vspace{0.5cm}
\caption{The time dependence of the von Neumann entropy (dashed lines),
Eq.(\ref{vonNeumannentropy}), and the loss of entanglement of formation $1 - E$ (solid lines),
given by Eq.(\ref{entanglementofformationforlambda}), are plotted for the experimental mean
value $\bar\lambda= 1.84 \cdot 10^{-12}\;\textrm{MeV}$ (lower curve) and the upper bound
$\bar\lambda_{up} = 4.34 \cdot 10^{-12}\;\textrm{MeV}$ (upper curve), Eq.(\ref{ergebnis}),
of the decoherence parameter $\lambda$. The time $t$ is scaled versus the lifetime $\tau_s$
of the short lived kaon $K_S$: $t \to t/\tau_s$. The vertical lines represent the propagation
time $t_0/\tau_s \approx 0.55$ of one kaon, including the experimental error bars, until it is
measured by the absorber in the CPLEAR experiment.}\label{entropyentanglementfigure}
\end{figure}

In Fig.\ref{entropyentanglementfigure} we have plotted the loss of entanglement $1 - E$,
given by Eq.(\ref{entanglementofformationforlambda}),
as compared to the loss of information, the von Neumann
entropy function $S$, Eq.(\ref{vonNeumannentropy}), in dependence of the time $t/\tau_s$
of the propagating \kkk system. The curves are shown for the mean value and upper bound of
the decoherence parameter $\lambda$, Eq.(\ref{ergebnis}). The von Neumann entropy function
visualizes the loss of the information about the correlation stored in the composite system.
Remember that the information flux is not flowing into the subsystems but into the
environment, see Eq.(\ref{entropysubsystems}). The loss of entanglement of formation
increases slower with time and visualizes the resources needed to create a given entangled
state. At $t=0$ the pure Bell state $\rho^{-}$ is created and becomes mixed for $t>0$ by
the other Bell state $\rho^{+}$. In the total state the amount of entanglement decreases
until separability is achieved (exponentially fast) for $t \to \infty$.

For example, in case of the CPLEAR experiment, where one kaon propagates about 2 cm,
which corresponds to a propagation time $t_0/\tau_s \approx 0.55$, until it is measured by
an absorber, the loss of entanglement is about $18\%$ for the mean value and maximal $38\%$
for the upper bound of the decoherence parameter $\lambda$. These values, however, could
diminish considerably in future experiments.

\section{Summary and conclusions}

We have considered a simple model of decoherence of the entangled \kkk state due to some
environment, i.e. a master equation of Liouville - von Neumann type with an additional
term $D[\rho]$. As generators of $D[\rho]$ causing the decoherence effect with strength
$\lambda$ we choose the projectors to the eigenstates of the ``effective mass''
Hamiltonian and, for simplicity, \cp invariance is assumed. For this choice the time
evolution for the components of $\rho$ decouples and only the off-diagonal elements are
effected by our modification.

We apply the model to the data of the CPLEAR experiment, where we follow the philosophy
that only the 2-particle state is affected by decoherence, whereas the 1-particle state
evolves according to pure QM. We estimate in this way the strength $\lambda$ of the
occurring decoherence, Eq.(\ref{ergebnis}).

Moreover, we can relate the model to the case of a phenomenologically introduced
decoherence parameter $\zeta$ and find a one-to-one correspondence, Eq.(\ref{zeta}).
However, the existing data are not yet sufficient to measure the time-dependence of $\zeta$
as predicted by our model, Eq.(\ref{zetamin}). So further measurements of the time-dependent
asymmetry term, Eqs.(\ref{lambdaasymmetry}), (\ref{zetaasymmetry}), would be of high
interest in future experiments and will sharpen considerably the bounds of
the parameters $\lambda$ and $\zeta$.

We can directly relate the decoherence of the \kkk state to its loss of entanglement. In
this connection we consider entanglement measures frequently discussed in the field of
quantum information. We demonstrate that the initially pure singlet state of the entangled
\kkk becomes mixed for $0<t<\infty$ but remains entangled and achieves separability for
$t \to \infty$ (see Proposition).

We find that entanglement loss in terms of the concurrence equals precisely the
decoherence parameter $\zeta$, Eq.(\ref{entanglementlossC}), and in terms of entanglement
of formation the loss is very well approximated by $\frac{\zeta}{\ln2}$ or
$\frac{\lambda}{\ln2}\cdot t$, Eq.(\ref{entanglementlossE}), which is one of our main
results. We can propose in this way how to measure experimentally the entanglement of the
\kkk system.

In Fig.\ref{entropyentanglementfigure} we visualize both the loss of information given by
the von Neumann entropy -- which flows totally into the environment and not into the
subsystems of the 2-particle system -- and the loss of entanglement of formation.
Inserting the mean value and upper bound of the parameter $\lambda$, which we have
determined from the CPLEAR experiment, we obtain definite bounds for both the
information- and entanglement loss of the propagating \kkk system in dependence of the
time. These values, however, could be improved considerably in future experiments.

\section{Acknowledgement}

The authors want to thank \v{C}aslav Brukner, Franz Embacher, Walter Grimus, Heide
Narnhofer and Walter Thirring for helpful discussions. This research has been supported
by the FWF Project No. P14143-PHY of the Austrian Science Foundation. The aid of the
Austrian-Czech Republic Scientific Collaboration, project KONTAKT 2001-11, and of the EU
project EURIDICE EEC-TMR program HPRN-CT-2002-00311 is acknowledged. Finally we would
like to thank our referees for helpful comments.

\begin{small}
\bibliographystyle{h-physrev}
\bibliography{revlind7a}
\end{small}

\end{document}